\documentclass[aps,prd,amsmath,twocolumn,superscriptaddress,preprintnumbers,amssymb,showpacs,floatfix,nofootinbib,longbibliography]{revtex4-1}
\pdfoutput=1

\usepackage{graphicx}
\usepackage{bm}
\usepackage{hyperref}
\usepackage{hyperref}
\usepackage{slashed}
\usepackage{aas_macros}
\usepackage{caption}
\usepackage{float}
\usepackage{slashed}
\usepackage{lipsum}
\usepackage{subfigure}
\usepackage{multirow}
\usepackage{amsmath}
\usepackage{array} 
\usepackage{varwidth} 
\usepackage{tabularx}
\usepackage{braket}
\usepackage[dvipsnames]{xcolor}

\newcommand{\be}{\begin{equation}}
\newcommand{\ee}{\end{equation}}
\newcommand{\bea}{\begin{eqnarray}}
\newcommand{\eea}{\end{eqnarray}}

\hypersetup{
     colorlinks   = true,
     citecolor    = teal,
     urlcolor     = teal,
     linkcolor    = teal
}

\usepackage{listings}
\usepackage{color,xcolor}

\begin{document}

\preprint{UCI-HEP-TR-2022-28}

\title{Bounds on Long-lived Dark Matter Mediators from Neutron Stars}

\author{Thong T.Q. Nguyen}
\thanks{{\scriptsize Email}: \href{mailto:ntqthonghep@gate.sinica.edu.tw}{ntqthonghep@gate.sinica.edu.tw}; {\scriptsize ORCID}: \href{https://orcid.org/0000-0002-8460-0219}{0000-0002-8460-0219}}
\affiliation{Institute of Physics, Academia Sinica, Nangang, Taipei 11529, Taiwan}
\affiliation{Sorbonne Université, 4 place Jussieu, F-75005 Paris, France}

\author{Tim M.P. Tait}
\thanks{{\scriptsize Email}: \href{mailto:ttait@uci.edu}{ttait@uci.edu}; {\scriptsize ORCID}: \href{https://orcid.org/0000-0003-3002-6909}{0000-0003-3002-6909}}
\affiliation{Department of Physics and Astronomy,  University of California, Irvine, CA 92697-4575 USA}

\begin{abstract}
Neutron stars close to the Galactic center are expected to swim in a dense background of dark matter.  For models in which the dark matter
has efficient interactions with neutrons, they are expected to accumulate their own local cloud of dark matter, making them appealing targets
for observations seeking signs of dark matter annihilation.  For theories with very light mediators, the dark matter may annihilate into pairs
of mediators which are sufficiently long-lived to escape the star and decay outside it into neutrinos.  We examine the bounds on the parameter
space of heavy ($\sim$~TeV to $\sim$~PeV)
dark matter theories with long-lived mediators decaying into neutrinos based on observations of high energy neutrino
observatories, and make projections for the future.  We find that these observations provide information that is complementary to
terrestrial searches, and probe otherwise inaccessible regimes of dark matter parameter space.
\end{abstract}

\maketitle

\section{Introduction}

The identity of the dark matter, a necessary ingredient to describe observations of the Universe \cite{Bertone:2004pz}, is a pressing question for particle physics which requires physics beyond the Standard Model (SM). There are a host of visions for how the SM may be supplemented with new particles and forces related to dark matter, and a rich program of experimental searches seeking to elucidate its nature \cite{Bertone:2018krk}.

One particular construction that has received a lot of recent attention posits that the dark matter interacts with the SM fields via a dark force carrier, which could either interact directly with the SM, or pick up interactions via mixing with the SM gauge or Higgs bosons.  When such particles are reasonably strongly interacting, such that they decay promptly when produced in terrestrial experiments, there are strong constraints from e.g. the Large Hadron Collider.  In the regime of smaller couplings to the SM, such particles can be very long-lived, and can be searched for using techniques to observe exotic long-lived particles \cite{Anchordoqui:2021ghd,FASER:2018eoc}. For even smaller couplings, the lifetime of the new force carriers may be too long for them to decay on planetary scales \cite{Feng:2015hja,Feng:2016ijc}, and constraints are typically somewhat weaker.   In this regime, production of the mediators from distant sources such as dark matter annihilation in extrasolar objects can provide key information through indirect searches for dark matter annihilation \cite{Slatyer:2017sev, Slatyer:2021qgc}.

Celestial bodies such as neutron stars located in dark matter-rich regions offer a particularly attractive observation target \cite{Drlica-Wagner:2022lbd}. Such objects can naturally sweep through a richer dark matter environment than is present in the solar system, if the dark matter has sufficiently strong scattering with them such that it can lose kinetic energy and become gravitationally bound, accumulating a local cloud of dark matter that is highly over-dense compared to the ambient density. The fate of this over-dense cloud of dark matter depends on its microscopic properties.  If it is sufficiently stable or has highly suppressed annihilation reactions, it can be considered an inert component of the celestial body itself. If it decays into SM particles or mediators with short lifetimes, it contributes a new heating process which could impact the energy budget of the celestial body~\cite{Goldman:1989nd,Kouvaris:2007ay,Bertone:2007ae,deLavallaz:2010wp,Kouvaris:2010vv, Guver:2012ba,McCullough:2010ai,Baryakhtar:2017dbj,Bell:2018pkk,Garani:2018kkd,Chen:2018ohx,Dasgupta:2019juq,Hamaguchi:2019oev,Camargo:2019wou,Bell:2019pyc, Garani:2019fpa,Acevedo:2019agu,Joglekar:2019vzy,Joglekar:2020liw,Bell:2020jou,Dasgupta:2020dik,Garani:2020wge,Bose:2022ola, Biswas:2022cyh, Bose:2021cou,Acuna:2022ouv,Shakeri:2022dwg, Leane:2020wob, Smirnov:2022tcg}. 
However, if it annihilates or decays into dark mediator particles whose lifetimes are sufficiently long as to allow them to escape before decaying, but nonetheless decay before reaching the Earth, it offers a unique opportunity to probe weakly-coupled dark mediators~\cite{Leane:2021ihh, Leane:2021tjj}. With their expected large Galactic populations in extremely dark-matter-rich environments, neutron stars could conceivably collectively power a bright DM annihilation signal.

\begin{figure}[t]
\centering
\captionsetup{justification=centering}
\includegraphics[width=0.8\columnwidth]{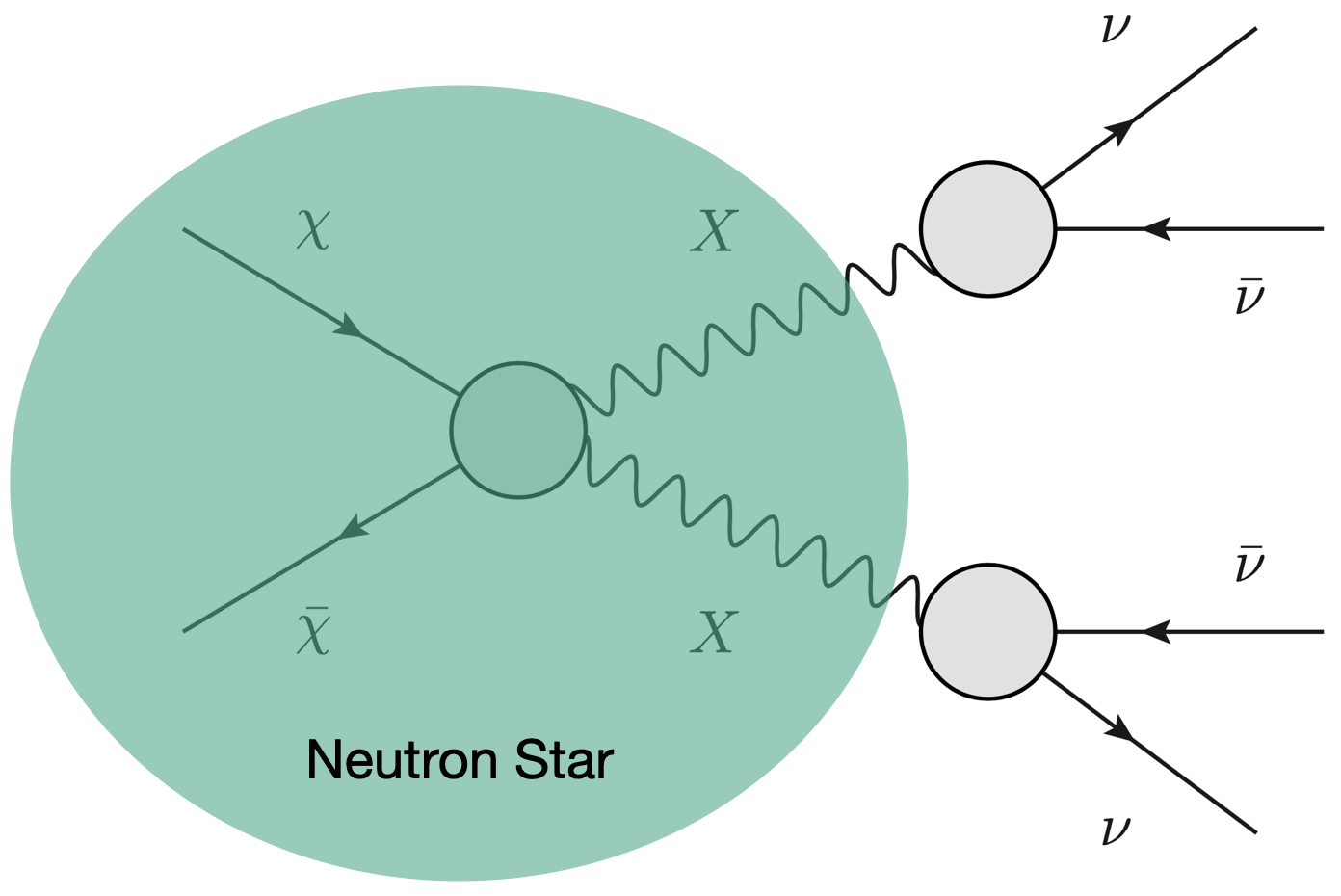}
\caption{Schematic diagram of the DM accumulated around a celestial body annihilating into long-lived mediator particles, which escape from the celestial body and decay into neutrinos.}
\label{fig:DPhotonescape}
\end{figure}

$\sim$~TeV photons and neutrinos are particularly promising messengers from dark matter annihilation, because they are relatively undeflected as they traverse Galactic distances
and thus can be traced back along the line from their origin. In this work, we consider the flux of high energy neutrinos that could be produced by heavy dark matter
particles (assumed for simplicity to be a Dirac fermion) which annihilates into a pair of spin-1 mediators which are long-lived, eventually decaying into neutrinos. 
Building on the work of Ref.~ \cite{Bose:2021yhz}, we contrast the predictions from this simplified model with measurements of the muon neutrino flux from $10^{3}$ to $10^{6}$ GeV from the Galactic center (GC) by IceCube \cite{IceCube:2017trr} and ANTARES \cite{ANTARES:2016mwq, ANTARES:2015moa} to place limits on the spin-independent cross-section, which controls the rate of accumulation and thus the density of the dark matter collected by the neutron star. We further consider the diffuse neutrino flux up to $10^{7}$ GeV measure by IceCube \cite{IceCube:2016umi, IceCube:2016qvd}, and find that it can provide important constraints for higher dark matter masses, and.  consider the reach of future high energy neutrino observatories such as ARIA \cite{Anker:2019rzo} to extend this sensitivity in the future.  

Our paper is organized as follow. In Sec. \ref{sect:DM_model}, we gave a brief review of the simplified model for dark matter under consideration. In Sec. \ref{sect:DMcapt}, we review dark matter capture and annihilation in celestial bodies, and derive experimental upper limits on the spin-independent cross section. Finally, in Sec. \ref{sect:conclusion}, we conclude with a summary of our main results and discuss possible future directions of our work.

\section{Dark Matter Simplified Model}
\label{sect:DM_model}

We consider a specific simplified model for the Dirac fermion dark matter $\chi$ which is a singlet under the SM gauge interactions, but interacting with 
a mediator corresponding to a $U(1)_{X}$ gauge symmetry. 
The dark gauge boson $X_\mu$ picks up interactions with the SM via kinetic mixing:
\begin{equation}
\begin{split}
\mathcal{L}\supset&  -\frac{1}{4}X_{\mu\nu}X^{\mu\nu}-\frac{\epsilon}{2}X_{\mu\nu}B^{\mu\nu}\\
&-\frac{1}{2}m_{X}^{2}X_{\mu}X^{\mu} + \bar{\chi}(i\slashed{D}_{U(1)_{X}}-m_{\chi}) \chi,
\end{split}
\label{eq:lagrangian}
\end{equation}
where $\epsilon$ parametrizes the kinetic mixing between $X_\mu$ and SM hypercharge interaction.
In general the SM fermions (and Higgs) transform under $U(1)_{X}$ as well,
and their interactions arise as a mixture of contributions from kinetic mixing, mass mixing, and 
direct couplings ~\cite{Carena:2004xs,Chen:2022abz, Bauer:2018onh}.

We consider DM above the unitarity limit, which typically requires a non-thermal production mechanism. We are interested in the limit in which $X$ is long-lived and decays primarily into neutrinos.  This naturally occurs in the limit where
$m_X$ is less than $2 m_e$, for which decays into other SM fermions are kinematically forbidden, 
and is further enhanced when the kinetic mixing approximately balances the direct couplings, such that the $X$ coupling to electrons
is suppressed compared to its coupling to neutrinos, which suppresses the one loop decay of $X$ into three photons and allows the decay into
neutrinos to dominate.

\begin{figure}[t]
\centering
\captionsetup{justification=centering}
\includegraphics[width=0.9\columnwidth]{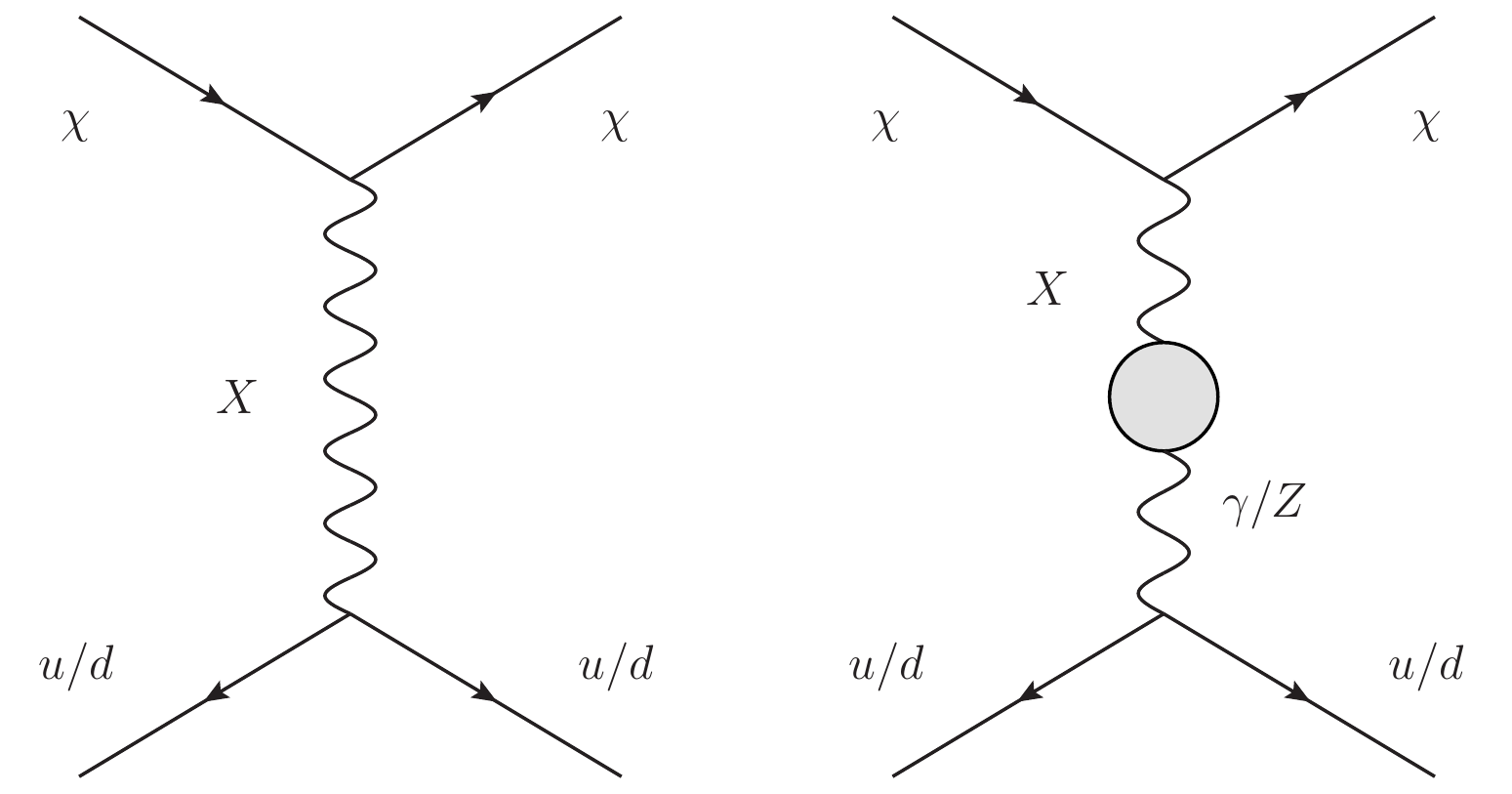}
\caption{Non-relativistic scattering between fermionic dark matter with quarks in nucleons (proton and neutron). Dark vector propagator either interacts with quarks directly or through mixing with photon and $Z$ boson.}
\label{fig:scattering_diagrams}
\end{figure}

Scattering between the dark matter and nucleons is mediated by $X$ exchange, as shown in Fig.~\ref{fig:scattering_diagrams}.  Generically, there are contributions from both
vector and axial interactions.  When the interactions with the SM are dominated by kinetic mixing, the axial current arises from mixing with the $Z$ boson, which is highly suppressed
for $m_X \ll M_Z$, as is generically the case when the $X$ boson is long-lived enough to escape from a celestial body.  Consequently, the spin-independent (SI) interaction dominates the scattering with nuclei.

\section{Accumulation and Annihilation}
\label{sect:DMcapt}

The cloud of dark matter surrounding a celestial body today (for our purposes, neutron stars in the Galactic Bulge) represents a balance between accumulation of the
dark matter from the environment, driven by the rate at which it scatters with and becomes gravitationally bound to the celestial body, and annihilation among the bound particles themselves.
We begin by reviewing some basic properties of neutron stars.

Neutron stars (NSs) are dense stellar objects composed almost entirely of degenerate neutrons,
with masses ranging from 1-1.5 $M_{\odot}$, and radii $R_{\mathrm{NS}}\simeq 10$ km.
The large gravitational field at the surface results in escape velocity $v_{\mathrm{esc}}\simeq 2\times 10^{5}$ km/s, and leads to significant blue-shifting of dark matter falling into the
gravitational well of $\chi \equiv 1-\sqrt{1-2G_{N}M/R}$ and $v_{\mathrm{esc}} \simeq \sqrt{2\chi}$.
We approximate the entire NS population
as having mass 1.5 $M_{\odot}$ and distributed in the bulge with a radial number density extrapolated from the ‘Fiducial $\times$ 10’ model of Ref.~\cite{Generozov:2018niv}:
\begin{align}
n_{\text{NS}}(r) &= \left\{
\begin{array}{lc}
5.98\times 10^{3}\Big{(}\frac{r}{1\text{pc}}\Big{)}^{-1.7}\text{pc}^{-3} &  ~(0.1 \text{pc}<r<2 \text{pc}) \\
2.08\times 10^{4}\Big{(}\frac{r}{1\text{pc}}\Big{)}^{-3.5}\text{pc}^{-3} &  (r>2 \text{pc}),
\end{array}
\right.
\label{eq:NS_population}
\end{align}

An interesting quantity characterizing a given celestial body is the saturation cross section, which is defined as
\begin{equation}
\sigma_{\mathrm{sat}} \equiv \pi R^{2}/N_{n},
\end{equation}
where $N_{n}$ is the total number of nucleons inside the celestial body. For neutron stars, this is
$\sigma^{{NS}}_{\mathrm{sat}}\sim 10^{-45}$ cm$^{2}$. 

\subsection{Dark Matter Capture}

As the celestial body moves through the dark matter halo, individual particles transit through it and can can scatter with its material and lose kinetic energy. 
After $N$-scatters, if its velocity falls below the escape velocity, it will be captured. 

\begin{figure}[t]
\centering
\captionsetup{justification=centering}
\includegraphics[width=\columnwidth]{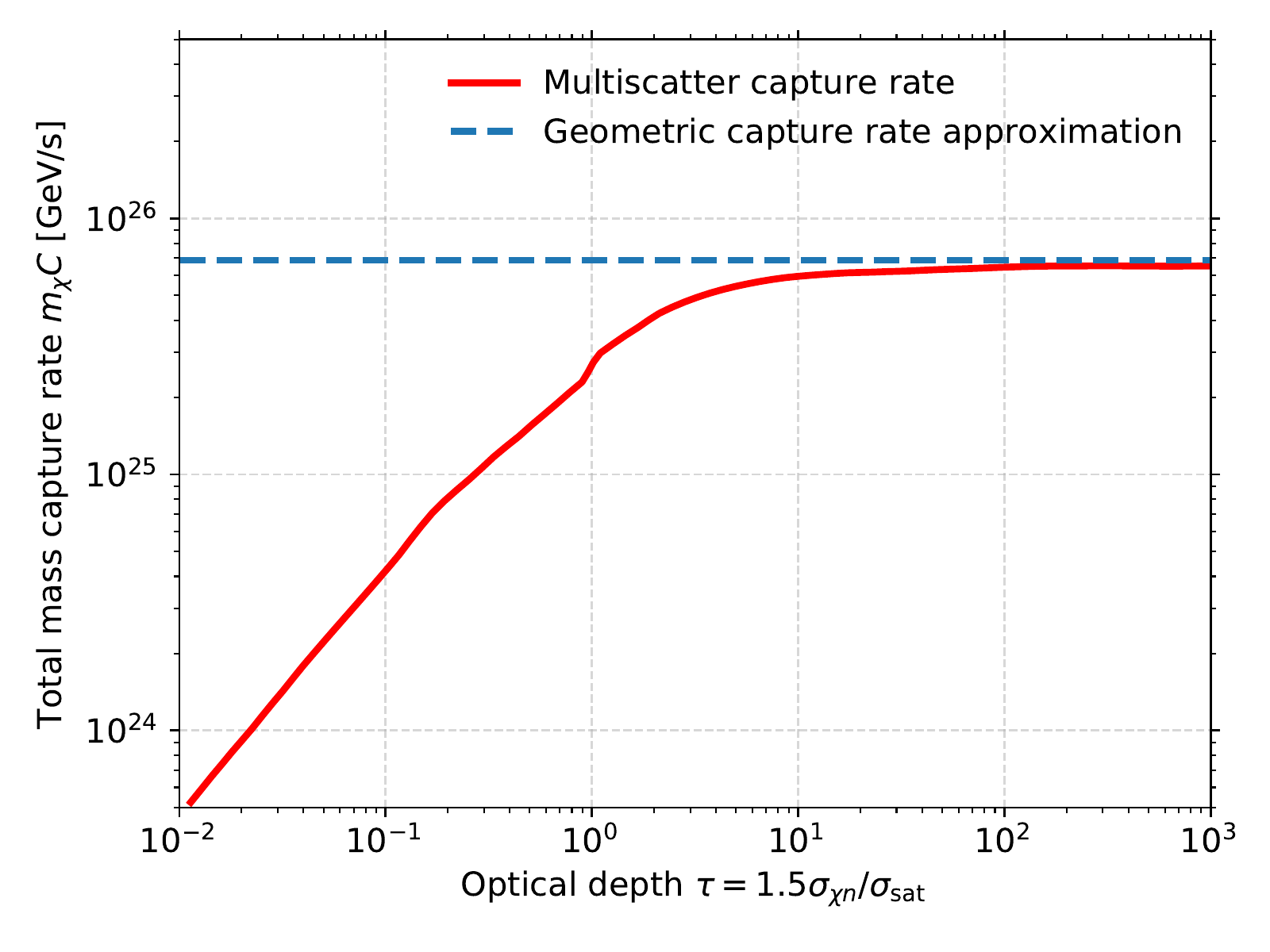}
\caption{The DM mass capture rate $m_{\chi}\times C$ as a function of the optical depth $\tau$ for a 
neutron star with $R=10$ km and $M=1.5 M_{\odot}$ and a dark matter model with 
$m_{\chi}=1$ TeV, $\rho_{\chi}=0.42$ GeV/cm$^{3}$, and $\bar{v}=220$ km/s.  The blue dashed line indicates $m_\chi C_{\rm max}$.\label{fig:SingleCapt}}
\end{figure}

The capture rate after scattering $N$ times is given by \cite{Leane:2021tjj}:
\begin{align}
&C_{N}(\tau) =\frac{\pi R^{2} ~p_{N}(\tau)}{(1-2G_{N}M/R)}\frac{\sqrt{6}n_{\chi}}{3\pi \bar{v}}\\
&\times\Big{[}(2\bar{v}^{2}+3v_{\text{esc}}^{2})-(2\bar{v}^{2}+3v_{N}^{2})\exp\Big{(}-\frac{3(v_{N}^{2}-v_{\text{esc}}^{2})}{2\bar{v}^{2}}\Big{)}\Big{]}\nonumber,
\label{eq:CaptN}
\end{align}
where $R$ is the radius of the celestial body,
$\bar{v}$ is the dispersion in the DM velocity distribution and $n_{\chi}$ is the local DM density.  $v_{N}$ is the typical DM velocity after scattering $N$ times,
taking into account the average energy lost in each scattering:
\begin{equation}
v_{N}=v_{\mathrm{esc}}\Big{(}1-\frac{\beta_{+}}{2}\Big{)}^{-N/2},
\end{equation}
where $\beta_{+}=4m_{\chi}m_{n}/(m_{\chi}+m_{n})^{2}$, and $p_{N}(\tau)$ is
the Poisson distribution for a given DM particle to undergo $N$ scatters:
\begin{equation}
p_{N}(\tau) = 2\int\limits_{0}^{1}dy\frac{ye^{-y\tau}(y\tau)^{N}}{N!},
\label{eq:probabilityN}
\end{equation}
which can be approximated in the limit of extreme single-scattering ($\tau\leq 1$) and multiple scattering ($\tau\gg 1$) \cite{Ilie:2020vec}.
The optical depth is defined as:
\begin{equation}
\tau = \frac{3}{2}\frac{\sigma_{\chi n}}{\sigma_{\mathrm{sat}}}.
\end{equation}
The total capture rate for an individual celestial object is:
\begin{equation}
C=\sum\limits_{N=1}^{\infty}C_{N},
\end{equation}
which in practice can be truncated at some large $N$.  Figure~\ref{fig:SingleCapt} shows the
rate of dark matter mass captured by a neutron star with $R=10$ km and $M=1.5 M_{\odot}$ 
as a function of the optical depth $\tau$, for a particular choice of dark matter with $m_{\chi}=1$ TeV, and halo described locally by
$\rho_{\chi}=0.42$ GeV/cm$^{3}$, and $\bar{v}=220$ km/s.

For large scattering cross sections, it approaches the maximum (``geometric" \cite{Garani:2017jcj}) capture rate $C_{N}\to p_{N}(\tau)\times C_{\mathrm{max}}$:
\begin{equation}
C_{\max}=\pi R^{2}n_{\chi}(r)v_{0}\Big{(}1+\frac{3}{2}\frac{v^{2}_{\text{esc}}}{\bar{v}(r)^{2}}\Big{)}\xi(v_{p}, \bar{v}(r)),
\label{eq:capture_max}
\end{equation}
where $v_{0}=\sqrt{8/3\pi}\bar{v}$ and $\xi(v_{p}, \bar{v}(r)) \sim 1$ takes into account the relative motion between the body and the DM halo.

The total capture rate in all of the objects within a spherical slice between radii $r_1$ and $r_2$ is
\begin{equation}
C_{\mathrm{total}}=4\pi\int\limits_{r_{1}}^{r_{2}}r^{2}n_{\star}(r)C dr,
\label{eq:Capt_all}
\end{equation}
where $n_{\star}(r)$ is the number density of the neutron stars [cf. Eqs.~(\ref{eq:NS_population})]. 
Choosing $r_{1}=0.1$ pc avoids poor modeling of the DM halo at the Galactic center, and $r_{2}=100$ pc reflects the fact that the populations of interest fall off rapidly away from
the center.

\subsection{Modeling the Galactic Center}
\label{ssect:MWGC}

\begin{table*}[t!]
\centering
\begin{tabular}{cccc}
\hline
Mass Component & Total mass ($M_{\odot}$) & Scale radius (kpc) & Center density ($M_{\odot}$ pc$^{-3}$)\\
\hline
Black hole & $4\times 10^{6}$ & --- & ---\\
Inner bulge (core) & $5\times 10^{7}$ & 0.0038 & $3.6\times 10^{4}$\\
Main Bulge & $8.4\times 10^{9}$ & 0.12 & $1.9\times 10^{2}$\\
Disk &  $4.4\times 10^{10}$ & 3.0 & 15\\
Dark halo & $5\times 10^{10}$ & $h=12.0$ & $\rho=0.011$\\
\hline
\end{tabular}
\caption{Parameters for exponential sphere model of the Galactic bulge, from \cite{Sofue:2013kja}.}
\label{tab:MWmass}
\end{table*}

The velocity dispersion of the dark matter at the location of each celestial object, $\bar{v}$, determines how much energy it must lose on the average through scattering in order to be captured, and thus plays an important role in the rate at which it accumulates dark matter.  At a give radius $r$ from the Galactic center, the
velocity dispersion is related to the circular velocity at that radius by $\bar{v}=\sqrt{3/2}v_{c}$, where $v_{c}$ depends on the total mass enclosed by a sphere of radius $r$:
\begin{equation}
v_{c}=\sqrt{\frac{G_{N}M(r)}{r}}.
\label{eq:vc}
\end{equation} 

The mass enclosed $M(r)$ can be modeled based on distinct components \cite{Sofue:2013kja}: the supermassive black hole Sag A*, an inner bulge, the main bulge, an exponential disk component, and the DM halo itself. Each component is described by the parameters shown in Table~\ref{tab:MWmass}, obtained from fits to stellar kinematics. The total mass is:
\begin{align}
    M(r) &= M_{\rm BH} \\
    &+4\pi \int_0^r r^{2} (\rho_{\rm outer} + \rho_{\rm inner} + \rho_{\rm disk} + \rho_{\rm DM}) dr\nonumber .
\label{eq:Milky_Way_Mass}
\end{align}
 We describe the DM density profile as a generalized NFW density profile \cite{Navarro:1995iw}:
\begin{equation}
 \rho_\chi(r)=\frac{\rho_0}{(r/r_s)^\gamma(1+(r/r_s))^{3-\gamma}},
 \label{eq:DM_density}
\end{equation}
where $r_{s}=12$ kpc is the scale radius, and $\rho_{0}=0.42 \text{ GeV/cm}^{3}$ is the local DM density.  We vary the inner slope index $\gamma=1.0-1.5$, as is motivated by expectations based on adiabatic contraction in the inner Galaxy in \cite{https://doi.org/10.48550/arxiv.1108.5736, DiCintio:2014xia}. As a cross-check, we compare our results to recent analyses of GC mass density profile in \cite{Cautun_2020, Foster:2021ngm} and find good agreement with the derived velocity dispersions.

\subsection{Annihilation to Long-lived Mediators}
\label{ssect:DM_ann}

As the cloud of DM accumulates around a celestial body, its constituents begin to annihilate.   
The evolution of their number $N(t)$ over time is governed by the equation \cite{Kouvaris:2010vv}:
\begin{equation}
\frac{dN(t)}{dt}=C_{\text{total}}-C_{A}N(t)^{2},
\label{eq:DM_number_evolve}
\end{equation}
where
\begin{equation}
C_{A}=\Braket{\sigma_{A}v}/V_{\text{eff}},
\label{eq:eff_ann_rate}
\end{equation}
is the annihilation rate, written in terms of the average annihilation cross section and the effective volume $V_{\text{eff}}$, approximated
as the volume of the celestial object: $V_{\text{eff}}=4\pi R^{3}/3$. The solution of Eq.~(\ref{eq:DM_number_evolve}) takes the form:
\begin{equation}
N(t) = \sqrt{\frac{C_{\text{total}}}{C_{A}}}\tanh\frac{t}{t_{\text{eq}}},
\label{eq:DM_number_solution}
\end{equation}
where 
\begin{equation}
t_{\rm eq} = 1/\sqrt{C_A C_{\rm tot}}
\label{eq:teq}
\end{equation}
is the timescale to attain equilibrium between DM capture and annihilation. 
The annihilation rate for non-self-conjugate DM at time $t$ is \cite{Leane:2021tjj}:
\begin{equation}
\Gamma_{\text{ann}}(t)=\frac{N(t)^{2}}{4V_{\text{eff}}}\Braket{\sigma_{A}v} .
\label{eq:AnnRate_General}
\end{equation}
We neglect the impact of DM evaporation, concentration diffusion, and thermal diffusion (for discussion, see Ref.~\cite{Leane:2022hkk}).

For $t \gg t_{\text{eq}}$, the number of DM particles becomes approximately time-independent, and its density inside the object reaches equilibrium. As discussed in Refs.~\cite{Acevedo:2019agu, Garani:2020wge}, the thermalization of DM is very rapid, even for tiny dark matter-nucleon scattering cross sections. Furthermore, neutron stars in the Galactic Center are expected to be very old ( $\gtrsim 5$ Gyr \cite{Generozov:2018niv}), such that the DM would reach equilibrium on a timescale much shorter than the age of a neutron star, and the annihilation rate simplifies to:
\begin{equation}
\Gamma_{\text{ann}} \rightarrow \frac{\Gamma_{\text{cap}}}{2}=\frac{C_{\text{total}}}{2} .
\label{eq:AnnRate_equi}
\end{equation} 

\begin{figure*}[t]
\centering
\captionsetup{justification=centering}
\includegraphics[width=1\columnwidth]{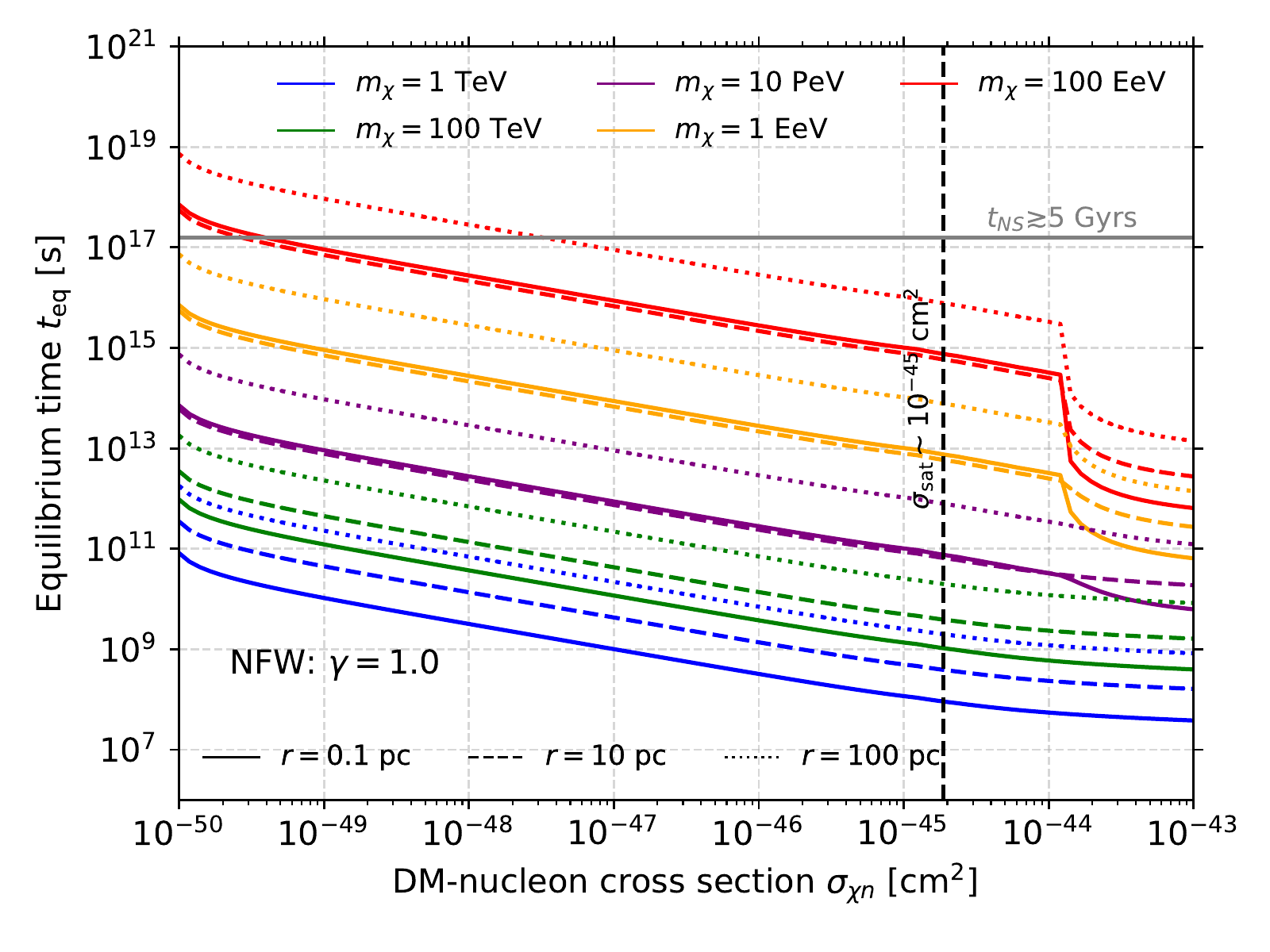}
\hfill
\includegraphics[width=1\columnwidth]{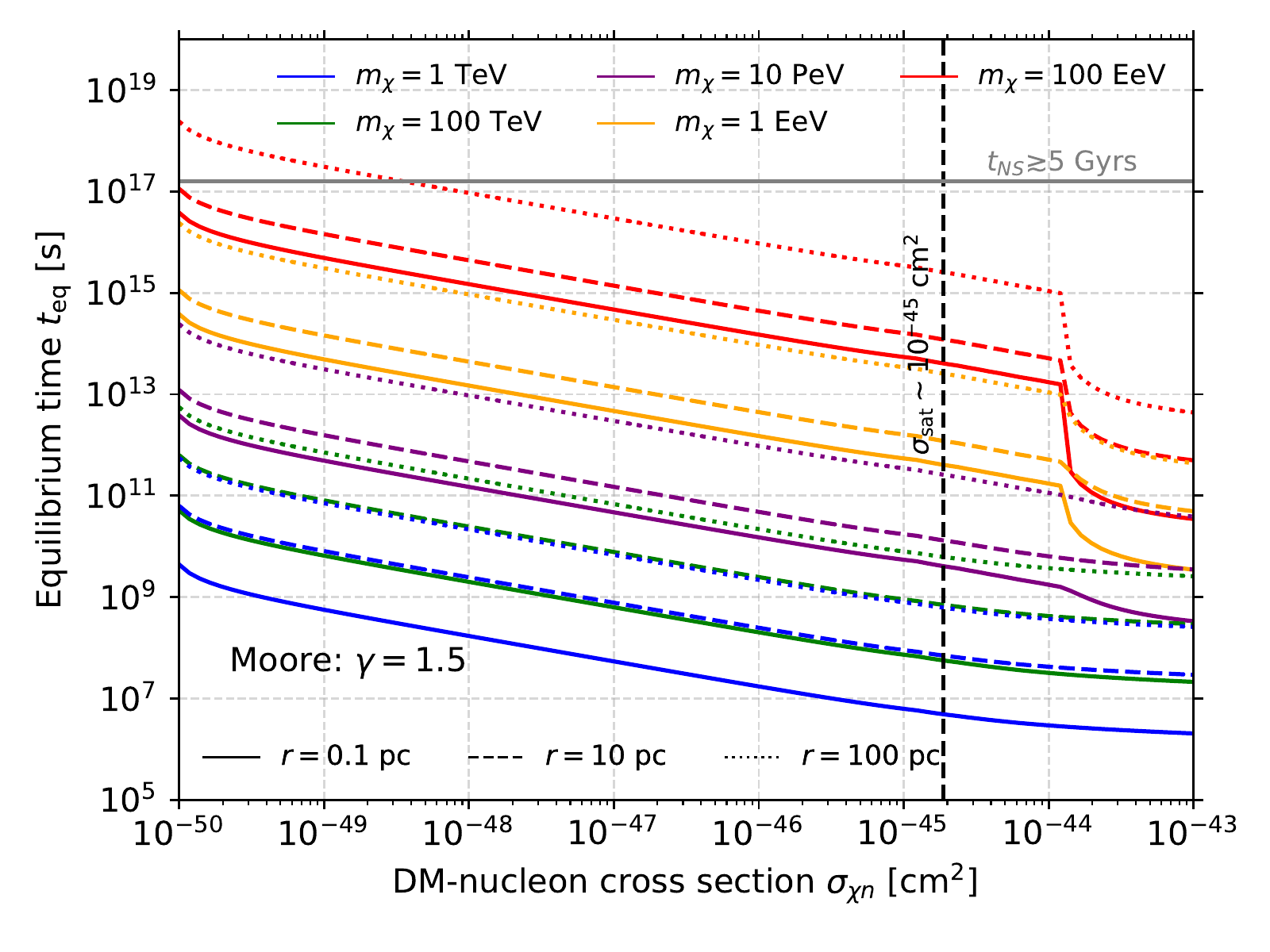}
\caption{Equilibrium timescales of Neutron Stars with $R=10$ km, $M=1.5M_{\odot}$ for different dark matter masses and at different positions in Galactic Center, as a function of DM-nucleon cross section. The thermal averaged annihilation cross section $\Braket{\sigma_{\chi\chi}v}\approx 3.3\times 10^{-26}$ cm$^{3}$/s is assumed. Bigger annihilation rates will lead to smaller equilibrium timescales, according to Equation \ref{eq:teq}. {\bf(Left)} NFW density profile: $\gamma=1.0$. {\bf(Right)} Moore density profile: $\gamma=1.5$.}
\label{fig:timescale}
\end{figure*}

\begin{figure*}[t]
\centering
\captionsetup{justification=centering}
\includegraphics[width=2\columnwidth]{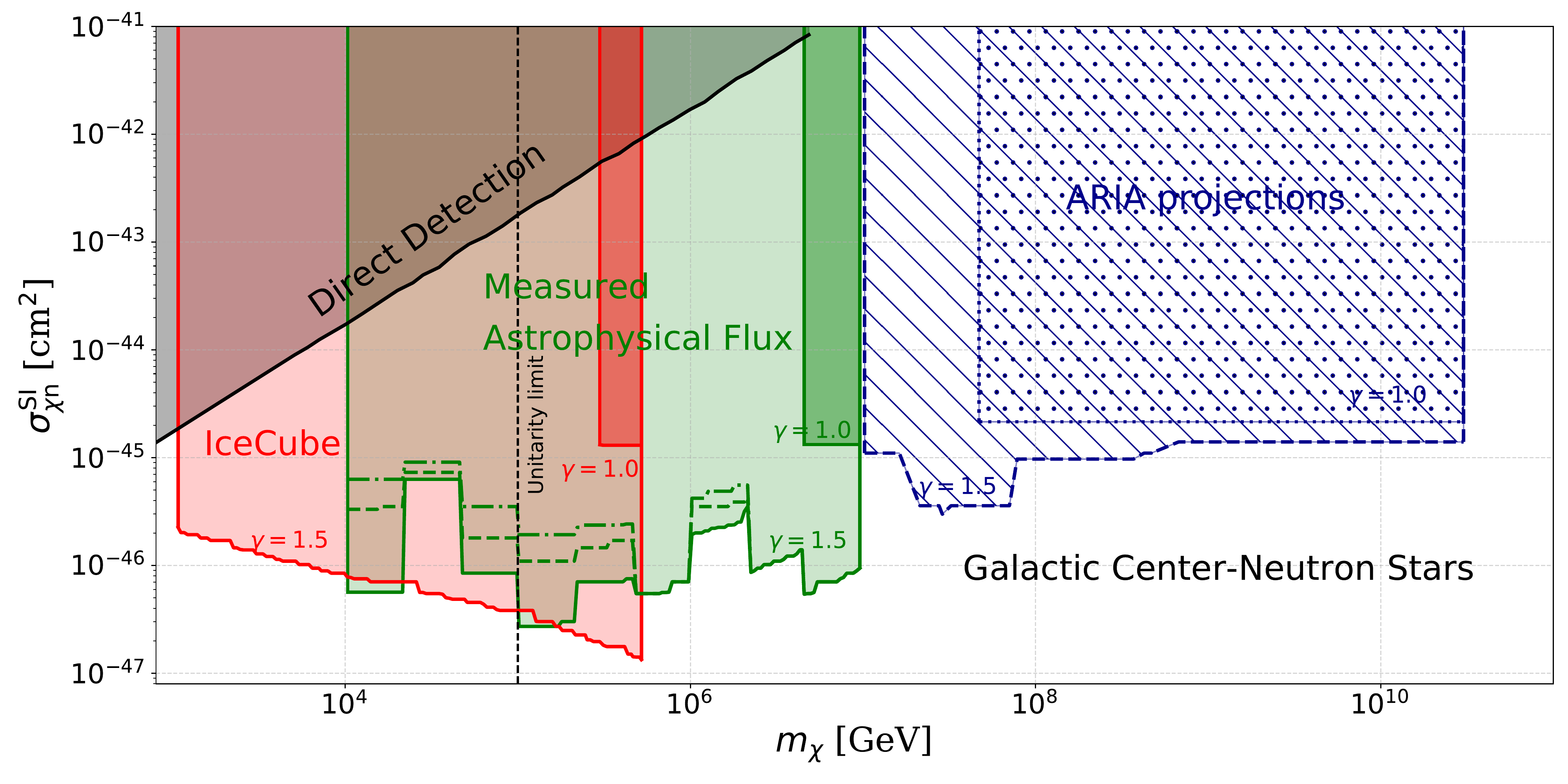}
\caption{Upper limits on the spin-independent cross section of dark matter scattering with neutrons, in the scenario in which accumulated dark matter annihilates into mediators which escape the neutron star and subsequently decay into neutrinos, for dark matter profiles described by $\gamma=1.0$ and $\gamma=1.5$.  The limits are derived from 
IceCube observations of the Galactic center (red) and the diffuse neutrino flux (green).  Projected limits from ARIA are shown in dark blue. Existing direct detection limits are indicated by the dark shaded region.}
\label{fig:xsectall}
\end{figure*}

In Figure \ref{fig:timescale}, we compare the equilibrium timescales for neutron stars at different positions in the GC with different DM masses. Assuming a thermal averaged annihilation cross section of $\Braket{\sigma_{\chi\chi} v}\approx 3.3 \times 10^{-26}$ cm$^{3}$/s, we observe that most of the parameter phase space in our considered DM mass range leads to the celestial objects reaching the equilibrium time scale. For small cross sections below $10^{-49}$ cm$^{2}$, the equilibrium breaks down for very large mass dark matter ($\sim$ EeV). However, as shown in Figure \ref{fig:xsectall} the DM reaches equilibrium at all of the relevent parameter space.

We consider the case in which the dark matter annihilates into a pair of mediators which are sufficiently weakly interacting with ordinary matter so as to be 
capable of escaping from the effective volume of the celestial body, after which they decay into high energy neutrinos (see Figure~\ref{fig:DPhotonescape}).  
Dark matter bound to the celestial body is expected to annihilate
with small relative velocity, resulting in the Lorentz boost factor of the produced mediators being approximately:
\begin{equation}
\eta \simeq m_{\chi}/m_{\phi},
\label{eq:Lorentz_boost}
\end{equation}
where $m_{\phi}$ is the mediator mass.

The differential energy flux ($\Phi$) of the neutrinos arriving to a detector on the Earth is given by \cite{Leane:2017vag}:
\begin{equation}
E^{2}\frac{d\Phi}{dE}=\frac{\Gamma_{\text{ann}}}{4\pi D^{2}} \times E^{2}\frac{dN}{dE}\times \text{BR}(X \to \nu \bar{\nu})\times P_{\text{surv}},
\label{eq:energy_flux}
\end{equation}
where $D$ is the average distance between the celestial object and the detector, and the $\mathrm{BR}(X \to \nu \bar{\nu})$ is the branching ratio for the mediator into neutrinos. 

\begin{figure}[t]
\centering
\captionsetup{justification=centering}
\includegraphics[width=\columnwidth]{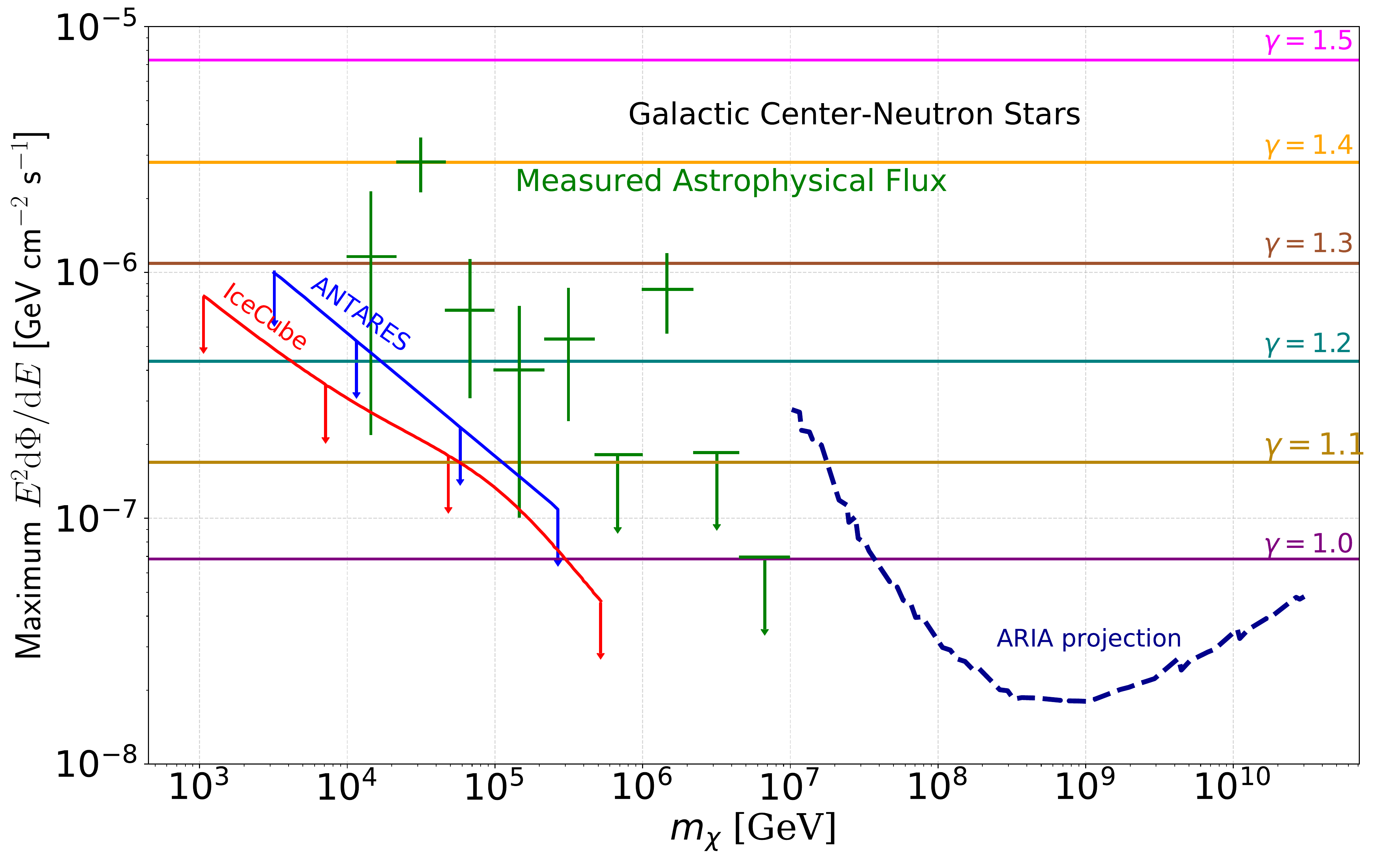}
\caption{Maximally optimistic values of differential flux of muon neutrinos from Galactic Center neutron stars. The IceCube 7-year upper limit is shown in red, the ANTARES limit in blue, and the measured IceCube diffuse flux in green. Future projected upper limits from ARIA are indicated in dark blue. }
\label{fig:maxcapt}
\end{figure}

For annihilation into a pair of mediators followed by 2-body mediator decay into neutrinos\footnote{More complex DM models \cite{Ramos:2021txu, Tran:2022yrh} require more complex modeling of the neutrino energy spectrum.}, 
their energy spectrum is a box distribution \cite{Ibarra:2012dw}:
\begin{equation}
\frac{dN_{\nu}}{dE_{\nu}} = \frac{4}{m_\chi} \Theta \left( E_\nu \right)  \Theta \left( m_\chi - E \right) ,
\end{equation}
where $\Theta (x)$ is the Heaviside step function.  $P_{\text{surv}}$ represents the probability that a mediator with lifetime $\tau_X$
decays outside of the celestial object and before reaching the detector located at a distance $D$ away:
\begin{equation}
P_{\text{surv}}=e^{-R/\eta c \tau_X} - e^{-D/\eta c \tau_X}.
\label{eq:surv_probability}
\end{equation}

\subsection{Bounds from Cosmic Neutrinos}
\label{ssect:bounds}

High energy neutrino observatories measuring the neutrino flux measure or place limits on $E^2 d\Phi/dE_\nu$, and thus (assuming equilibrium) on
the rate at which dark matter is accumulated on neutron stars, controlled by the microphysics of the dark matter scattering.  In particular, IceCube \cite{IceCube:2017trr}, and ANTARES
\cite{ANTARES:2016mwq} have upper limits on the muon neutrino flux from the direction of the Galactic Center 
(defined by $-40^{\circ}<l<40^{\circ}$ and $-3^{\circ}<b<3^{\circ}$ in Galactic coordinates), and IceCube \cite{IceCube:2015gsk} has measured the all-sky diffuse high
energy muon neutrino flux from TeV up to PeV energies.  In the future, the proposed ARIA experiment is expected to be able to measure neutrinos up to $10^{11}$~GeV~\cite{Anker:2019rzo,ARIANNA:2019scz}.  We assume that the mediator decays democratically into all three flavors of active neutrino. Given the large distances involved from production to the detector, we further assume that neutrino flavor oscillations wash out such that the signal incident on the Earth is well-described by a flavor ratio of $1:1:1$.
Under the maximally optimistic assumption that the capture rate is large enough to saturate at $C_{\max}$ and $P_{\text{surv}} = 1$, we show the largest possible signal for different choices of the dark matter inner slope profile
$\gamma$ are shown in Figure \ref{fig:maxcapt}, with the flux limits and measurements overlaid for context.  For an NFW profile ($\gamma=1$), one could expect to see features from
dark matter producing neutrinos for a narrow range of dark matter masses.  For steeper profiles, the impact of dark matter annihilations becomes pronounced.

Moving away from the maximally optimistic scenario, we derive limits on the cross section for spin-independent scattering of dark matter with
neutrons, as a function of the dark matter mass and in the limit in which $m_X \ll m_\chi$ and for mediator lifetimes 
$R \ll \eta c \tau_X \ll D$.   We make the conservative assumption 
that the entire neutrino flux arises from the dark matter signal, ignoring contributions e.g. from supernovae \cite{Westernacher-Schneider:2019utn, Brdar:2016ifs}.  Realistic modeling of backgrounds could reduce the potential contribution from dark matter annihilation, and lead
to stronger limits.  We compare with the IceCube observations of the Galactic center (which are strongest for $m_\chi \lesssim 10^6$~GeV)
and measurements/limits of the diffuse flux (which are important for $10^6$~GeV $\lesssim m_\chi \lesssim 10^7$~GeV) and with projected
flux limits from ARIA, which provide key information for dark matter masses $\gtrsim 10^7$~GeV.
In Figure~\ref{fig:xsectall}, we show the resulting limits and projected limits (for dark matter profiles with $\gamma=1, 1.5$), 
together with existing limits from terrestrial dark matter
searches \cite{PICO:2019vsc, PandaX-4T:2021bab}, for comparison.  These results make it clear that existing high energy neutrino
observatories provide unique information about 
theories of dark matter with light long-lived mediators decaying into neutrinos, and that future proposals such as ARIA can dramatically
extend this information to probe new regimes of dark matter parameter space.

\section{Conclusions}
\label{sect:conclusion}

In this paper, we have examined the prospects to use neutrinos produced by the decay of long-lived mediators, themselves produced by
the annihilation of the cloud of dark matter that is expected to accumulate around neutron stars in the Galaxy.  These types of theories are
notoriously difficult to probe with terrestrial searches, and our results (summarized in Figure~\ref{fig:xsectall})
highlight the important contributions of existing high energy neutrino observatories, and the potential for future projects \cite{Ackermann:2022rqc} such
as ARIANNA, ARIA, IceCube-Gen2 \cite{IceCube-Gen2:2020qha, Ishihara:2019aao}, KM3Net \cite{KM3Net:2016zxf}, ANITA-IV \cite{ANITA:2019wyx}, PUEO \cite{PUEO:2020bnn}, RNO-G \cite{RNO-G:2020rmc}, and Auger \cite{PierreAuger:2019ens} to probe
otherwise inaccessible territory of dark matter model-space.

\section*{Acknowledgment}
We thank Payel Mukhopadhyay, Rebecca Leane, Felix Kahlhoefer, Stephanie Wissel,
Chau Thien Nhan, Nicholas Rodd, Meng-Ru Wu, Yen-Hsun Lin, Tzu-Chiang Yuan and Martin Spinrath for fruitful discussions. We also thank the anonymous referee for the constructive comments and suggestions. The work of T.T.Q.N. as a research assistant is supported in part by the Ministry of Science and Technology (MOST) of Taiwan under Grant No 111-2112-M-001-035 and by the Sciences and Engineering Faculty of Sorbonne Université. T.T.Q.N. would like to thank the Department of Physics and Center for Theory and Computation, NTHU, Taiwan for its hospitality. The work of T.M.P.T. is supported in part by the US National Science Foundation under Grant PHY-2210283.

\bibliography{biblio.bib}
\end{document}